\documentclass[12pt]{article}
\usepackage{graphicx}
\usepackage{rotating}
\usepackage{amsmath}

\newcommand{\BABARPubYear}    {05}
\newcommand{\BABARPubNumber}  {112}
\newcommand{\SLACPubNumber} {11491}

\RequirePackage{xspace}

\usepackage{relsize}
\def\babar{\mbox{\slshape B\kern-0.1em{\smaller A}\kern-0.1em
    B\kern-0.1em{\smaller A\kern-0.2em R}}}

\def\Kbar  {\kern 0.2em\overline{\kern -0.2em K}{}\xspace}

\def\Kz    {\ensuremath{K^0}\xspace}
\def\Kzb   {\ensuremath{\Kbar^0}\xspace}
\def\KzKzb {\ensuremath{\Kz \kern -0.16em \Kzb}\xspace}
\def\Kp    {\ensuremath{K^+}\xspace}
\def\Km    {\ensuremath{K^-}\xspace}

\def\KpKm  {\ensuremath{\Kp \kern -0.16em \Km}\xspace}

\def\Dbar    {\kern 0.2em\overline{\kern -0.2em D}{}\xspace}

\def\Dz      {\ensuremath{D^0}\xspace}
\def\Dzb     {\ensuremath{\Dbar^0}\xspace}
\def\DzDzb   {\ensuremath{\Dz {\kern -0.16em \Dzb}}\xspace}
\def\Dp      {\ensuremath{D^+}\xspace}
\def\Dm      {\ensuremath{D^-}\xspace}

\def\DpDm    {\ensuremath{\Dp {\kern -0.16em \Dm}}\xspace}

\def\Bbar    {\kern 0.18em\overline{\kern -0.18em B}{}\xspace}

\def\Bz      {\ensuremath{B^0}\xspace}
\def\Bzb     {\ensuremath{\Bbar^0}\xspace}
\def\BzBzb   {\ensuremath{\Bz {\kern -0.16em \Bzb}}\xspace}
\def\Bu      {\ensuremath{B^+}\xspace}
\def\Bub     {\ensuremath{B^-}\xspace}

\def\BpBm    {\ensuremath{\Bu {\kern -0.16em \Bub}}\xspace}

\def\BorBbar    {\kern 0.18em\optbar{\kern -0.18em B}{}\xspace}
\def\DorDbar    {\kern 0.18em\optbar{\kern -0.18em D}{}\xspace}
\def\KorKbar    {\kern 0.18em\optbar{\kern -0.18em K}{}\xspace}

\mathchardef\Upsilon="7107
\def\Y#1S{\ensuremath{\Upsilon{(#1S)}}\xspace}

\mathchardef\Deltares="7101
\mathchardef\Xi="7104
\mathchardef\Lambda="7103
\mathchardef\Sigma="7106
\mathchardef\Omega="710A

\def\Deltabar{\kern 0.25em\overline{\kern -0.25em \Deltares}{}\xspace}
\def\Lbar{\kern 0.2em\overline{\kern -0.2em\Lambda\kern 0.05em}\kern-0.05em{}\xspace}
\def\Sigbar{\kern 0.2em\overline{\kern -0.2em \Sigma}{}\xspace}
\def\Xibar{\kern 0.2em\overline{\kern -0.2em \Xi}{}\xspace}
\def\Obar{\kern 0.2em\overline{\kern -0.2em \Omega}{}\xspace}
\def\Nbar{\kern 0.2em\overline{\kern -0.2em N}{}\xspace}
\def\Xb{\kern 0.2em\overline{\kern -0.2em X}{}\xspace}

\newcommand{\tev}{\ensuremath{\mathrm{\,Te\kern -0.1em V}}\xspace}
\newcommand{\gev}{\ensuremath{\mathrm{\,Ge\kern -0.1em V}}\xspace}
\newcommand{\mev}{\ensuremath{\mathrm{\,Me\kern -0.1em V}}\xspace}
\newcommand{\kev}{\ensuremath{\mathrm{\,ke\kern -0.1em V}}\xspace}
\newcommand{\ev}{\ensuremath{\mathrm{\,e\kern -0.1em V}}\xspace}
\newcommand{\gevc}{\ensuremath{{\mathrm{\,Ge\kern -0.1em V\!/}c}}\xspace}
\newcommand{\mevc}{\ensuremath{{\mathrm{\,Me\kern -0.1em V\!/}c}}\xspace}
\newcommand{\gevcc}{\ensuremath{{\mathrm{\,Ge\kern -0.1em V\!/}c^2}}\xspace}
\newcommand{\mevcc}{\ensuremath{{\mathrm{\,Me\kern -0.1em V\!/}c^2}}\xspace}

\def\mus  {\ensuremath{\rm \,\mus}\xspace}

\def\mus        {\ensuremath{\,\mu{\rm s}}\xspace}    

\def\ra                 {\ensuremath{\rightarrow}\xspace}

\def\pep2{PEP-II}

\def\gsim{{~\raise.15em\hbox{$>$}\kern-.85em
          \lower.35em\hbox{$\sim$}~}\xspace}
\def\lsim{{~\raise.15em\hbox{$<$}\kern-.85em
          \lower.35em\hbox{$\sim$}~}\xspace}

\xspace

\newcommand{\jprlBase}       {Phys.\ Rev.\ Lett.\xspace}
\newcommand{\jprBase}        {Phys.\ Rev.\xspace}
\newcommand{\jplBase}        {Phys.\ Lett.\xspace}
\newcommand{\plb}       [1]  {\jplBase\ B~{\bf #1}}

\newcommand{\jprl}      [1]  {\jprlBase\ {\bf #1}}
\newcommand{\jprd}      [1]  {\jprBase\ D~{\bf #1}}

\def\jetset74   {\mbox{\tt Jetset \hspace{-0.5em}7.\hspace{-0.2em}4}\xspace}

\def\Y#1S{{\Upsilon\rm(#1S)}}

\def\ra{\rightarrow}

\def\beq{\begin{equation}}
\def\eeq{\end{equation}}

\begin{document}

\begin{flushright}
SLAC-PUB-\SLACPubNumber \\
\babar-TALK-\BABARPubYear/\BABARPubNumber\\
September 2005
\end{flushright}

\par\vskip 4cm

\begin{center}
\Large \bf New Hadron Spectroscopy with \babar $^*$
\end{center}
\bigskip

\begin{center}
Gagan B. Mohanty\\
{\em Representing the \babar\ Collaboration}\\
Department of Physics, University of Warwick, Coventry CV4 7AL, UK
\end{center}

\bigskip \bigskip

\begin{abstract}
We review hadron spectroscopy at \babar\ with emphasis on recent
results from the studies of the $X$(3872) state, inclusive charmonia on
recoil, double charmonium production, and the broad structure observed
at around 4.26 GeV/$c^2$. These results are preliminary, unless
otherwise specified.
\end{abstract}

\vfill
\begin{center}
Presented at Frontier Science 2005, New Frontier in Subnuclear Physics,\\
Milan, Italy, 12-17 September 2005.\\
Submitted to Frascati Physics Series (Proceedings)
\end{center}

\vspace{1.0cm}
\begin{center}
{\em Stanford Linear Accelerator Center, Stanford University, 
Stanford, CA 94309} \\ \vspace{0.1cm}\hrule\vspace{0.1cm}
$^*$Work supported in part by Department of Energy contract DE-AC02-76SF00515.
\end{center}

\newpage

\section{Introduction}
Hadron spectroscopy plays a crucial role in validating quantum
chromodynamics (QCD) and the quark substructure of matter. In
recent years this field of particle physics has received a renewed
interest, thanks to the discovery of many new states at the $B$
factories. Some of these states, such as $D_{sJ}(2317)^+$ and
$D_{sJ}(2460)^+$, are by now well established and appear to be
ordinary charm mesons. Others, such as $X$(3872) and $Y$(3940),
could be new excited charmonium ($c\bar{c}$) states or new forms
of matter, but require more measurements with better statistical
precision for a definitive identification.

The \babar\ experiment \cite{babar} is an $e^+e^-$ collider experiment
taking data at or just below the $\Upsilon(4S)$ resonance at the PEP-II
asymmetric $B$ factory. It was designed to study CP violation in the
$B$ system, however it has proved to have a significantly broader
physics reach, especially in the spectroscopy sector. We report here
recent results on new charmonium-like states studied using the \babar\
data sample.
\section{Studies of the $X$(3872) State}
The $X$(3872) state was first observed in the decay $B^-\ra X(3872)K^-,
X\ra J/\psi\pi^+\pi^-$ \cite{charge} by the Belle Collaboration \cite{belleX}.
Later it was confirmed by the CDF \cite{cdfX}, D0 \cite{d0X} and \babar\
\cite{babarX1} Collaborations. The distribution of the $\pi^+\pi^-$ invariant
mass in the decay $X(3872) \ra J/\psi\pi^+\pi^-$ suggests that the decay may
proceed through an intermediate $\rho^0$ resonance. If so, one can expect to
find its charged isospin partner $X(3872)^-$. We have searched for this
state in the decays $B^- \ra J/\psi\pi^-\pi^0K^0_S$ and $B^0 \ra J/\psi\pi^-\pi^0
K^+$. However, no evidence for a charged $X(3872)$ has been found \cite{babarX2}
and we set the following upper limits at 90\,\% confidence level (CL):
$\mathcal{B}(B^- \ra X(3872)^-K^0_S, X^- \ra J/\psi\pi^-\pi^0) < 11 \times 10^{-6}$ and
$\mathcal{B}(B^0 \ra X(3872)^-K^+, X^- \ra J/\psi\pi^-\pi^0) < 5.4 \times 10^{-6}$.
A search for $X(3872)$ in the process $e^+e^- \ra X(3872)\gamma_{ISR}$ via the
decay to $J/\psi\pi^+\pi^-$, where $\gamma_{ISR}$ denotes initial state radiation
(ISR), has yielded a null result \cite{babarX3}. This strongly disfavors a
$J^{PC} = 1^{--}$ assignment to $X$. Coupled with the fact that $X(3872)$ has
a mass close to the $D^0\bar{D}^{0\star}$ threshold and has a very narrow width,
this seems to tell us that X(3872) is not a simple charmonium state.
Alternative proposals have been made to explain this state; for example,
it could be a weakly bound $D\bar{D^{\star}}$ molecule-like state \cite{theory1}
\begin{figure}
\begin{center}
\includegraphics[width=.49\columnwidth]{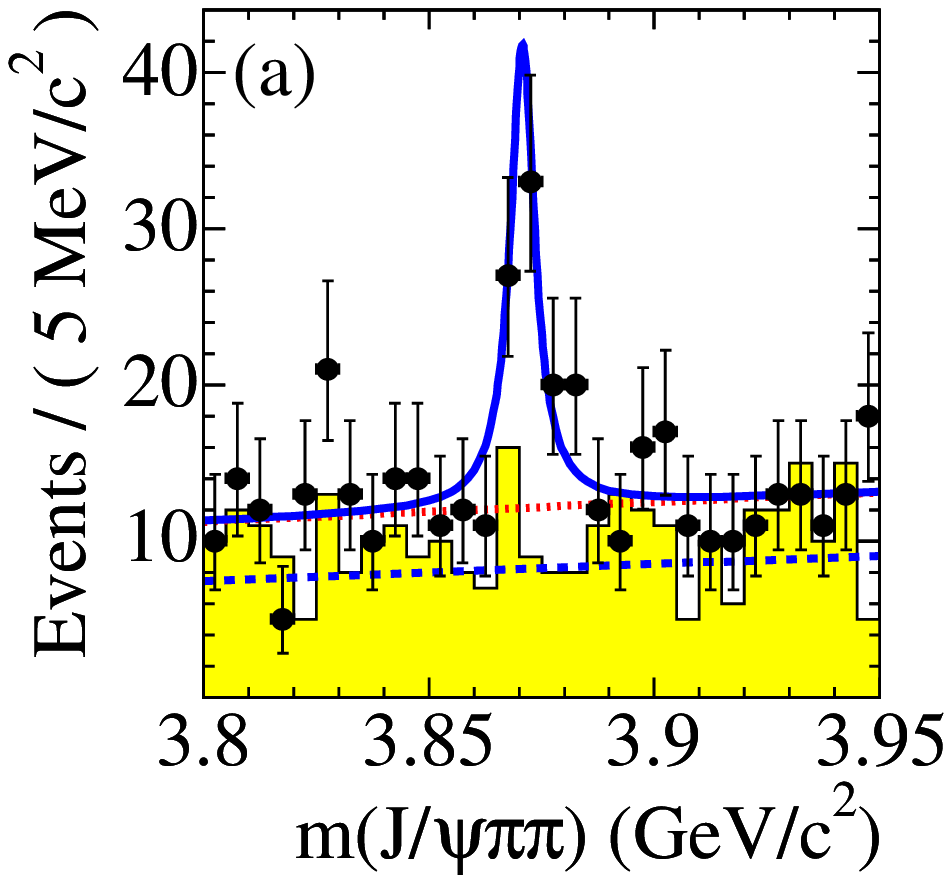}
\includegraphics[width=.49\columnwidth]{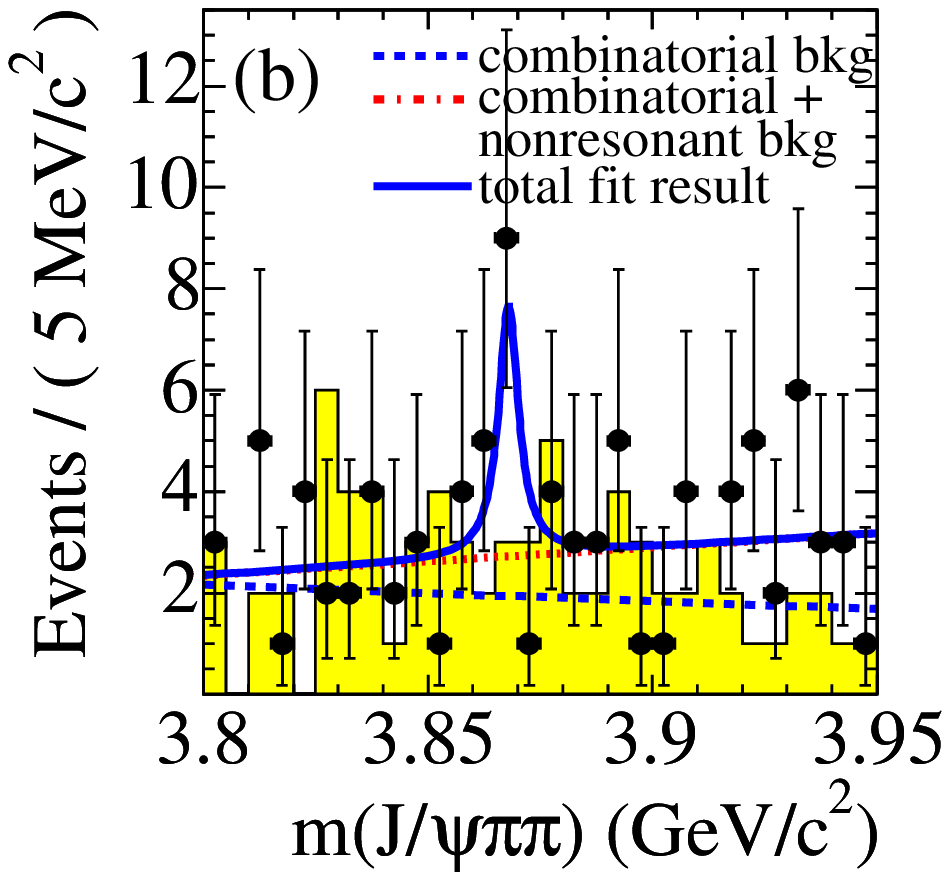}
\end{center}
\caption{$J/\psi\pi^+\pi^-$ invariant mass distribution in the
         signal region for (a) $B^- \ra X(3872)K^-$ and (b)
         $B^0 \ra X(3872)K^0_S$.}
\label{fig1}
\end{figure} 
or a diquark-antidiquark state \cite{theory2}. To investigate these models,
\babar\ has updated its earlier study with increased statistics.

Using 232 million $B\bar{B}$ pairs recorded by \babar, $X(3872)$ candidates are
reconstructed in the decays $B^- \ra J/\psi\pi^+\pi^-K^-$ and $B^0 \ra J/\psi
\pi^+\pi^-K^0_S$. Figure~\ref{fig1} shows the $J/\psi\pi^+\pi^-$ invariant mass
distributions for these two $B$ decay modes \cite{babarX4}. For the charged $B$
mode, we obtain $61.2\pm 15.3$ while for the $B^0$ mode only $8.3\pm 4.5$ signal
events. These yields are translated to the respective branching fractions by
taking efficiency corrections into account: $\mathcal{B}^- \equiv \mathcal{B}
(B^- \ra X(3872)K^-, X \ra J/\psi\pi^+\pi^-) = (10.1\pm 2.5\pm 1.0)\times 10^{-6}$
and $\mathcal{B}^0 \equiv \mathcal{B}(B^0 \ra X(3872)K^0_S, X \ra J/\psi\pi^+\pi^-)
= (5.1 \pm 2.8\pm 0.7)\times 10^{-6}$, where uncertainties are statistical and
systematic, respectively. From these we derive a ratio of the branching
fractions, $\mathcal{R} = \mathcal{B}^0/\mathcal{B}^- = 0.50\pm 0.30\pm 0.05$.
We also measure the mass difference of the $X(3872)$ state from the charged
and neutral $B$ decay modes, $\Delta m$, to be $(2.7\pm 1.3\pm 0.2)\,{\rm MeV}/c^2$.
The diquark-antidiquark model predicts $\mathcal{R} = 1$ and $\Delta m$ to be
$(7\pm 2)\,{\rm MeV}/c^2$. The expected ratio of branching fractions is consistent
with our measurement, $0.13 < \mathcal{R} < 1.10$ at 90\,\% CL and the observed
$\Delta m$ is both consistent with zero and the model prediction within two standard
deviations ($\sigma$). This result seems to slightly disfavor the molecule model
that predicts $\mathcal{R}$ to be at most 10\,\%. However, we need more data to
convincingly discriminate between these models.
\section{Inclusive Charmonia on recoil}
A novel recoil technique is devised to measure various charmonium states
$X_{c\bar{c}}$ in inclusive $B$ decays to the two-body final state $B^- \ra
K^-X_{c\bar{c}}$. Two body decays are identified by their characteristic
monochromatic line without any necessity of reconstructing the decay
products of $X_{c\bar{c}}$. Therefore, this method allows us to determine
the absolute branching fractions (or set upper limits) for the
production of any known or unknown charmonium resonances. The analysis
is performed using 210 ${\rm fb}^{-1}$ of data where a charged $B$
meson is fully reconstructed, so that the momentum of the recoiling $B$
candidate can be calculated from the measured $B$ and the beam parameters.
\begin{figure}
\begin{center}
\includegraphics[width=.49\columnwidth]{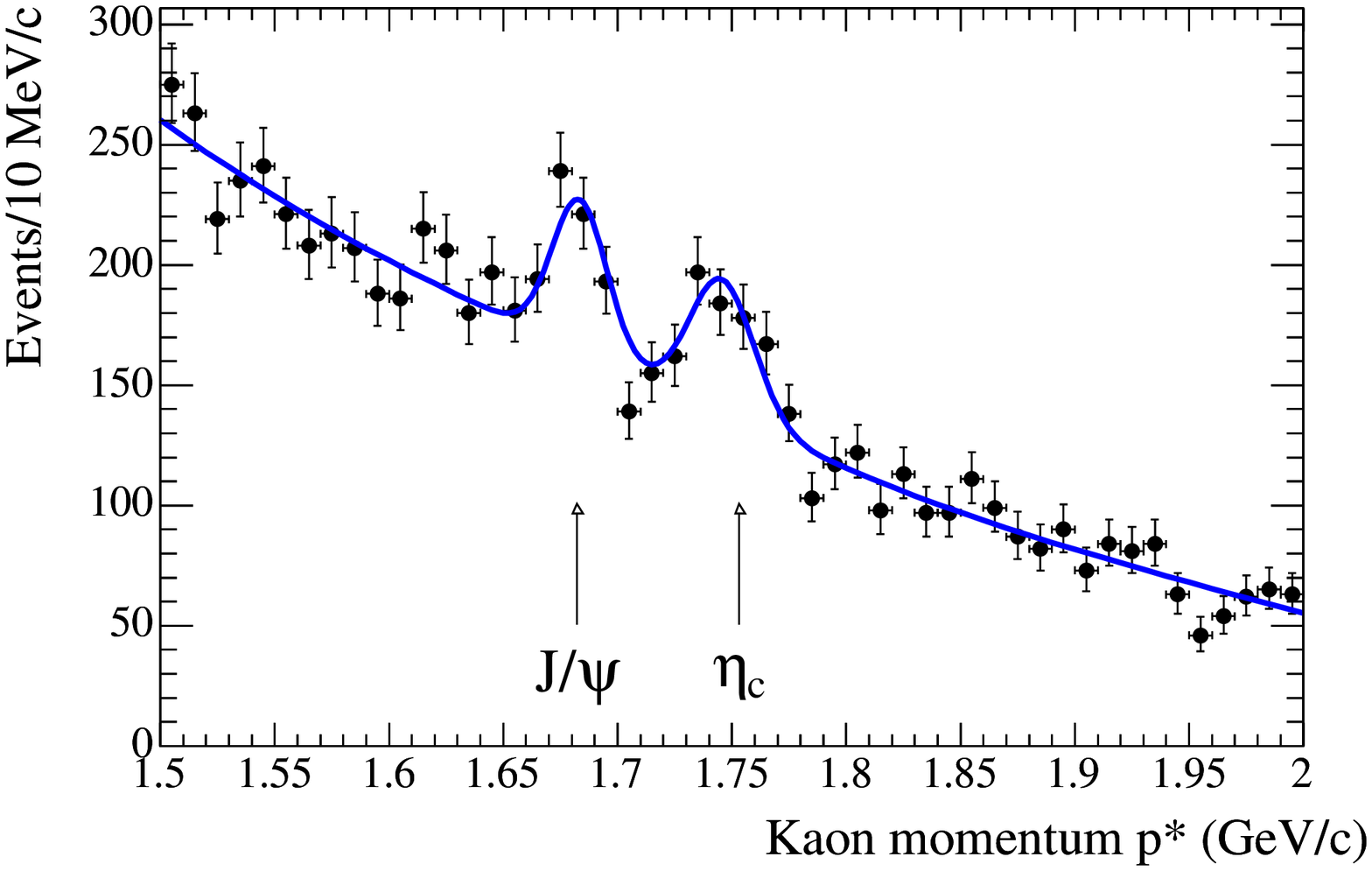}
\includegraphics[width=.49\columnwidth]{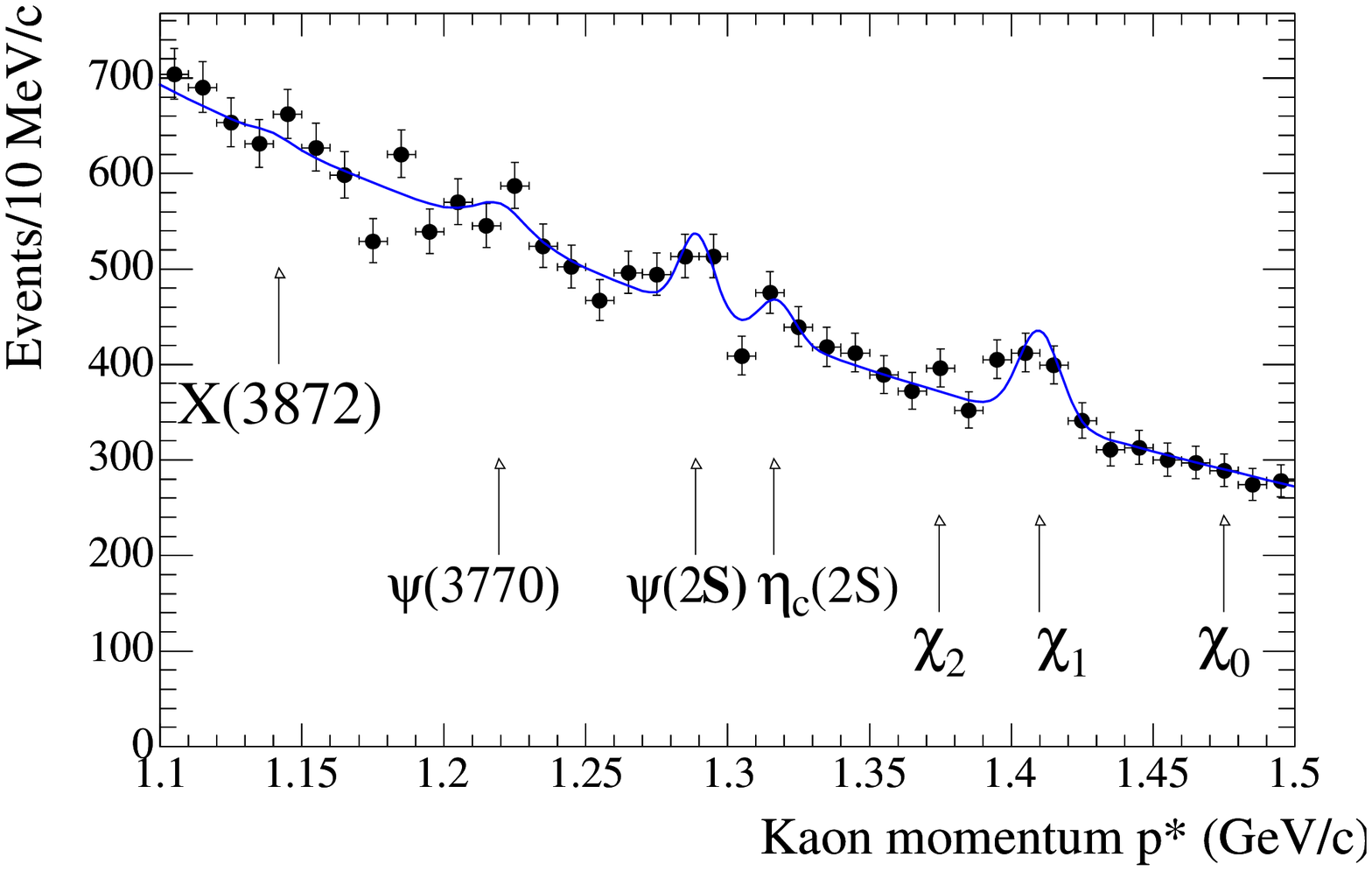}
\end{center}
\caption{Kaon momentum in the recoiling (second) $B$ meson rest frame.}
\label{fig2}
\end{figure} 
The kaon energy ($E^\star_K$) in the second $B$ rest frame is related to
the mass ($m_{\rm X}$) of the system recoiling against it by $m_{\rm X} =
\sqrt{m^2_B+m^2_K-2E^\star_K m_B}$, where $m_B$ is the charged $B$ meson
mass. Therefore, the signature of any charmonium states will be visible as
a peak in the kaon momentum (or energy) spectrum as shown in Figure
\ref{fig2}. The $J/\psi$, $\eta_c$ and $\chi_{c1}$ signals are clearly
seen in the figure. We also have 3.2\,$\sigma$, 1.8\,$\sigma$ and
1.4\,$\sigma$ indications for $\psi(2S)$, $\eta_c(2S)$ and $\psi(3770)$,
respectively. No evidence for $X(3872)$ is found and we derive $\mathcal{B}
(B^-\ra X(3872)K^-) < 3.2\times 10^{-4}$, which, in conjunction with the
world-average value of the branching fraction product,
allows us to set the lower limit $\mathcal{B}(X(3872)\ra J/\psi\pi^+\pi^-) >$
4.2\,\% at 90\,\% CL.
\section{Double Charmonium Production}
\begin{figure}
\begin{center}
\includegraphics[width=.6\columnwidth]{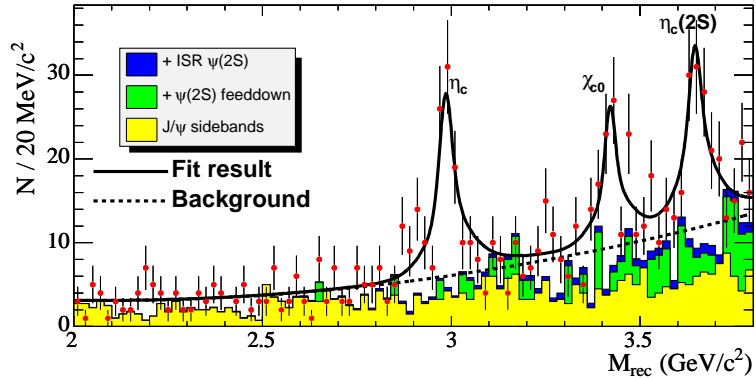}
\end{center}
\caption{Mass of the system recoiling against the reconstructed $J/\psi$.}
 \label{fig3}
\end{figure}

In \babar, exclusive $B$ decays are not the only source of charmonium
states. We have also studied double charmonium production in the process
$e^+e^-\ra \gamma^\star\ra J/\psi\,{c\bar{c}}$ using 124 ${\rm fb}^{-1}$
of data \cite{babarCC}. Only ${c\bar{c}}$ states with even C-parity are
expected in this reaction, although if there is a contribution from
$e^+e^-\ra \gamma^\star\gamma^\star\ra J/\psi\,{c\bar{c}}$, odd C-parity
states could also be produced. In this study, we first reconstruct a $J/\psi$
candidate via its leptonic decay mode and then calculate the mass of the
system recoiling against it (see Figure~\ref{fig3}). Three even C-parity
charmonium states, $\eta_c$, $\chi_{c0}$ and $\eta_c(2S)$, are observed,
while there is no evidence for any odd C-parity states such as $J/\psi$.
The distribution is fit to obtain the yield for each state, from which the
production cross section is calculated. Due to the requirement of at least
five charged tracks in the event for background suppression, we report the
product of the branching fraction to states with more than two tracks and
the production cross section. The results are $17.6\pm 2.8^{+1.5}_{-2.1}$ fb,
$10.3\pm 2.5^{+1.4}_{-1.8}$ fb and $16.4\pm 3.7^{+2.4}_{-3.0}$ fb for $\eta_c$,
$\chi_{c0}$ and $\eta_c(2S)$, respectively. These values are an order of magnitude
higher than those predicted by non-relativistic QCD \cite{theory3}.
However, recent works incorporating charm quark dynamics \cite{theory4}
seem to narrow down the discrepancy.
\section{Observation of $Y$(4260)}
ISR events produced in the $\Upsilon(4S)$ energy region at the $B$ factories
act as a probe of interesting physics occurring at a lower center-of-mass
energy. Motivated by this, \babar\ has investigated the $e^+e^-\ra J/\psi\pi^+
\pi^-\gamma_{ISR}$ process across the charmonium mass range, using a data sample
\begin{figure}
\begin{center}
\includegraphics[width=.6\columnwidth]{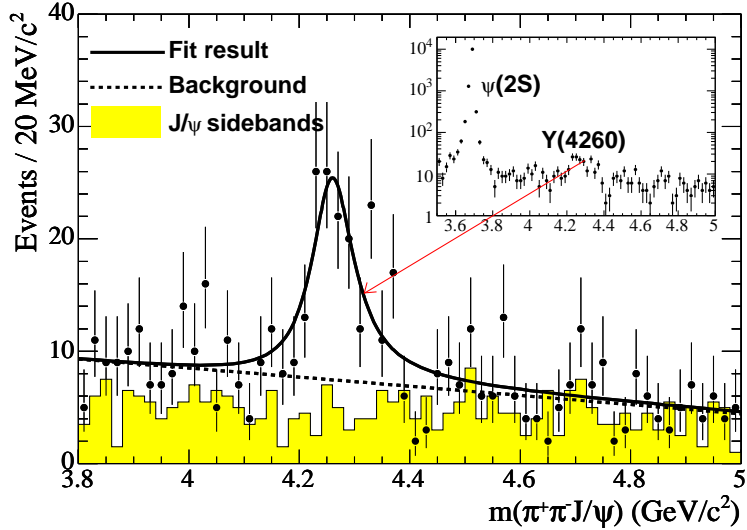}
\end{center}
\caption{$J/\psi\pi^+\pi^-$ invariant mass spectrum in the range 3.8$-$5.0
         ${\rm GeV}/c^2$ and (inset) over a wider range that includes the
         $\psi(2S)$ state.}
 \label{fig4}
\end{figure} 
of 233 ${\rm fb}^{-1}$ integrated luminosity \cite{babarY}. These events are
characterized by two pions, two leptons (electron or muon) making a $J/\psi$
candidate and a very small recoil mass against the $J/\psi\pi^+\pi^-$ system.
Figure~\ref{fig4} shows the $J/\psi\pi^+\pi^-$ invariant mass spectrum for the
selected candidates. An enhancement of events near 4.26 ${\rm GeV}/c^2$ is
clearly observed in addition to the expected $\psi(2S)$ peak. No other
structures are evident in the spectrum including the $X(3872)$. Using a maximum
likelihood fit, we obtain a signal yield of $125\pm 23$ with a statistical
significance of 8\,$\sigma$ (the signal is referred to as $Y(4260)$). The mass
and width of the particle are found to be $4259\pm 8^{+2}_{-6}\, {\rm MeV}/c^2$
and $88\pm 23^{+6}_{-4}$ MeV, respectively. We also calculate a value of
$\Gamma(Y(4260)\ra e^+e^-)\cdot\mathcal{B}(Y\ra J/\psi\pi^+\pi^-)
= 5.5\pm 1.0^{+0.8}_{-0.7}$ eV. Although all these results are from a single
resonance fit, we cannot exclude or establish a multi-resonance hypothesis at
the current level of statistics. More data are needed to reveal its exact nature.
\section{Conclusions}
The last few years have been very exciting for hadron spectroscopy studies at
the $B$ factories. Specifically, \babar\ is pioneering several sensitive
searches for new charmonium states, some of which are summarized in this paper,
including the first measurement of the $X(3872)$ state in neutral and charged
$B$ decays and the observation of a new broad resonance $Y(4260)$.


\begin{thebibliography}{99}
\bibitem{charge}  Charge conjugates are included implicitly throughout.
\end{thebibliography}
\end{document}